\documentstyle[preprint,aps]{revtex}  
\tightenlines
\title{Stability of critical behaviour of weakly disordered systems
with respect to the replica symmetry breaking}

\begin{document}
\draft
\author{V.V. Prudnikov, P.V. Prudnikov, and A.A. Fedorenko}
\address{Department of Theoretical Physics, Omsk State University,
Mira prospekt 55-A,  Omsk, 644077, Russia}
\date{January 25, 2001}
\maketitle
\begin{abstract}
A field-theoretic description of the critical behaviour of the weakly
disordered systems  is given.
Directly, for three- and two-dimensional systems a renormalization analysis
of the effective Hamiltonian of model with replica symmetry
breaking (RSB) potentials is carried out in the two-loop approximation.
For case with 1-step RSB the fixed points (FP's) corresponding to
stability of the various types of critical behaviour are identified
with the use of the Pade-Borel summation technique. Analysis of FP's has
shown a stability of the critical behaviour of the weakly disordered systems
with respect to RSB effects and realization of former scenario of disorder
influence on critical behaviour.
{\sloppy

}
\end{abstract}


The effects produced by weak quenched disorder on the critical phenomena
have been studied since many years ago \cite{1,2,3,4,5}. According to the
Harris criterion \cite{1}, the disorder effects the critical behaviour only
if $\alpha$, the specific heat exponent of the pure system, is positive.
In this case a new universal critical behaviour, with new critical exponents,
was established. In contrast, when $\alpha < 0$, the disorder appears to be
irrelevant for the critical behaviour.

In dealing with the weak quenched disorder the traditional approach is
the replica method \cite{4,5}, and in terms of replicas all the results
obtained for the disorder systems correspond to the so-called
replica-symmetric (RS) solutions. Physically it means that only unique ground
state is assumed to be relevant for the observable thermodynamics. However,
in a number of papers \cite{6,7,8} the ideas about replica symmetry breaking
(RSB) in the systems with quenched disorder were presented. For the first
time in the paper \cite{6} physical arguments  showing that
in the presence of the quenched disorder there exist numerous local minimal
energy configurations separated by finite barriers and demonstrating how the
summation over these local minima configurations can provide additional RSB
interaction potentials for fluctuating fields have been offered. The
renormalization group (RG)
description of the classical $\phi^4$ model with RSB potentials was
presented in the one-loop approximation using $\varepsilon$ - expansion
\cite{6,7,8}. It was shown that the RSB degrees of freedom produce dramatic
effect on the asymptotic behaviour of the RG flows, such that for a general
type of the RSB there exist no stable fixed points, and RG equations arrive
into the strong coupling regime. In contrast, in Ref.~\cite{9} 
using of the fermion representation it was demonstraited that the critical 
behaviour of the 2D weakly disordered Ising system is stable with respect to 
the RSB modes.

However, our numerous investigations of pure and disordered systems
performed in the two-loop and higher orders of the approximation
for the 3D system directly together with methods of series summation show
that the predictions made in the lowest order of the approximation,
especially on the basis of the $\varepsilon$ - expansion, can differ
strongly from the real critical behaviour \cite{10}. Therefore, the results
of RSB effects investigation in Refs.~\cite{6,7,8} must be reconsidered 
with the use of a more accurate field-theory approach in the higher orders of
the approximation.

In this paper, we have realized the field-theoretical RG description
in the two-loop approximation of the 3D and 2D models of the weakly
disordered systems with RSB interaction potentials of forth order on
fluctuating fields.  We have carried out the solution of the RG equations
with the use of series summation method and the analysis of stability of
various types of critical behaviour with respect to RSB effects.

We consider an $O(p)$ - symmetric Ginzburg-Landau-Wilson model of a spin
system with weak quenched disorder near critical point given by the
Hamiltonian
\begin{equation} \label{Ham_1}
 H=\int d^dx\left\{\frac{1}{2}\sum_{i=1}^{p}[\nabla{\phi}_{i}(x)]^{2}+
\frac{1}{2}[\tau-\delta\tau(x)]\sum_{i=1}^{p}{\phi}_{i}^{2}(x)+
\frac{1}{4}g\sum_{i,j=1}^{p}{\phi}_{i}^{2}(x){\phi}_{j}^{2}(x)\right\},
\end{equation}
where $\phi_{i}(x)$ is the $p$-component order parameter and
$\delta\tau(x)$ is the Gaussian-distributed random transition temperature
with the second moment of distribution
$\left<\!\left<(\delta\tau(x))^{2}\right>\!\right>\sim u$
defined by the positive constant $u$ which is proportional to the
concentration of defects.
The use of the standard replica trick gives the possibility to average
easily over the disorder and reduce the task of statistical description of
the weakly disordered system with the Hamiltonian (\ref{Ham_1}) to the
homogeneous system with the effective Hamiltonian
\begin{equation}  \label{Ham_2}
 H_n=\int d^dx\left\{\frac{1}{2}\sum_{i=1}^{p}\sum_{a=1}^{n}\left[
  [\nabla{\phi}_{i}^{a}(x)]^{2}+\tau[{\phi}_{i}^{a}(x)]^{2}\right]+
\frac{1}{4}\sum_{i,j=1}^{p}\sum_{a,b=1}^{n}g_{ab}[{\phi}_{i}^{a}(x)]^{2}
[{\phi}_{j}^{b}(x)]^{2}\right\},
\end{equation}
which is a functional of $n$ replications of the original order parameter
with an additional vertex $u$ in the replica symmetric matrix
$g_{ab}=g\delta_{ab}-u$. The properties of the original disordered system
are obtained in the replica number limit $n\rightarrow 0$. The following
standard RG procedure based on the statistical calculation of contribution
to the partition function of longwavelength order parameter fluctuations
around the global minimum state with $\phi(x)=0$ gives the possibility to
find the various types of critical behaviour and conditions of their
stability and carry out the calculation of critical exponents.

However, as it was shown in Refs.~\cite{6,7,8}, the fluctuations of
random transition temperature $\delta\tau(x)$ for $[\tau - \delta\tau(x)]<0$
can lead to realization in system of numerous regions with $\phi(x) \neq 0$
displaying through the numerous local minimal energy configurations separated
from the ground state by finite barriers. In this case the direct
application of the traditional replica-symmetric RG scheme may be
questioned. For statistical description of such systems near phase transition
point the Parisi RSB scheme was used in Refs.~\cite{6,7,8} by analogy with 
spin glasses \cite{11}. It was argued that spontaneous RSB can occur due to the
interaction of the fluctuating fields with the local non-perturbative
degrees of freedom coming from the multiple local minima solutions of the
mean-field equations. It was shown that the summation over these solutions
in the replica partition function can provide the additional non-trivial RSB
potential $\sum_{a,b}g_{ab}{\phi}_{a}^{2}{\phi}_{b}^{2}$ in which the
matrix $g_{ab}$ has the Parisi RSB structure \cite{11}. According to the
technique of the Parisi RSB algebra, in the limit $n\rightarrow 0$ the
matrix $g_{ab}$ is parametrized in terms of its diagonal elements $\tilde{g}$
and the off-diagonal function $g(x)$ defined in the interval $0<x<1$:
$g_{ab}\rightarrow (\tilde{g},g(x))$. The operations with the matrices
$g_{ab}$ are given by the following rules:
\begin{equation}  \label{g_ab}
 g_{ab}^{k}\rightarrow (\tilde{g}^{k};g^{k}(x)),
(\hat{g}^{2})_{ab}=\sum_{c=1}^{n}g_{ac}g_{cb}\rightarrow (\tilde{c};c(x)),
(\hat{g}^{3})_{ab}=\sum_{c,d=1}^{n}g_{ac}g_{cd}g_{db}\rightarrow
(\tilde{d};d(x)),
\end{equation}
where
\begin{equation}
 \tilde{c}= \tilde{g}^{2} - \int_{0}^{1}d x g^{2}(x), \qquad
    c(x) = 2 \left[\tilde{g} - \int_{0}^{1}d y g(y)\right]g(x) -
    \int_{0}^{x}d y [g(x) - g(y)]^2, \nonumber
\end{equation}
\begin{eqnarray} \label{c_d}
 \tilde{d}= \tilde{c}\tilde{g} - \int_{0}^{1}d x c(x)g(x), \qquad
    d(x) = \left[\tilde{g} - \int_{0}^{1}d y g(y)\right]c(x) +
    \left[\tilde{c} - \int_{0}^{1}d y c(y)\right]g(x) \nonumber \\
    - \int_{0}^{x}d y [g(x) - g(y)][c(x) - c(y)].
\end{eqnarray}
The RS situation corresponds to the case $g(x) = {\rm const}$ independent of
$x$.

We carried out the field-theoretical RG description of the 3D and 2D models
with the effective replicated Hamiltonian (\ref{Ham_2}) in which the matrix
$g_{ab}$ has the RSB structure in the two-loop approximation. In the
field-theoretic approach the asymptotic critical behavior of systems in
the fluctuation region are determined by the Callan-Symanzik RG equation
for the vertex parts of the irreducible Green's functions. To calculate
the $\beta$ functions as functions of the renormalized elements of
the matrix $g_{ab}$ appearing in the RG equation, we used the method based
on the Feynmann diagram technique and the renormalization procedure \cite{12}.
We obtained the next expressions for two-pont vertex function
$\Gamma^{(2)}$ and four-point vertex functions $\Gamma^{(4)}_{ab}$:
\begin{eqnarray}
 &&\left.\frac{\partial\Gamma^{(2)}}{\partial k^2}\right|_{k^2=0}=
    1+4fg_{aa}^2+2pf\sum\limits_{c=1}^{n}g_{ac}g_{ca}, \\
 &&\left.\Gamma^{(4)}_{ab}\right|_{k_i=0}=g_{ab}-p\sum\limits_{c=1}^
    {n}g_{ac}g_{cb}-4g_{aa}g_{ab}
    -4g_{ab}^2+(8+16h)g_{ab}^3+(24+8h)g_{aa}^2g_{ab} \nonumber\\
 &&\ \ \ \ \ +48hg_{aa}g_{ab}^2+4g_{aa}g_{bb}g_{ab}+8ph\sum
    \limits_{c=1}^{n}
    g_{ac}g_{cb}^2+8phg_{ab}\sum\limits_{c=1}^{n}g_{ac}g_{cb}   \nonumber  \\
 &&\ \ \ \ \ +4phg_{ab}\sum\limits_{c=1}^{n}g_{ac}^2+2p\sum
    \limits_{c=1}^{n}g_{ac}
    g_{cc}g_{cb}+4pg_{aa}\sum\limits_{c=1}^{n}g_{ac}g_{cb}+p^2\sum
    \limits_{c,d=1}^{n}g_{ac}g_{cd}g_{db},
\end{eqnarray}
where
\begin{eqnarray}
&&\left. f(d)=-\frac{1}{J^2}\frac{\partial}{\partial k^2}\int
  \frac{d^{d}k_1d^{d}k_2}{(k_1^2+1)(k_2^2+1)((k_1+k_2+k)^2+1)}
  \right|_{k^2=0}, \nonumber \\
&&h(d)=\frac{1}{J^2}\int\frac{d^{d}k_1d^{d}k_2}{(k_1^2+1)(k_2^2+1)
  ((k_1+k_2)^2+1)},  \\
&&J = \int d^{d}k/(k^2+1)^2, \  \ f(d=3)=\frac{2}{27}, \  \ h(d=3)=\frac{2}{3}, \nonumber \\
&&f(d=2)=0.11464,  \  \ h(d=2)=0.78129, \nonumber
\end{eqnarray}
and was made the redefinition $g_{ab}\rightarrow g_{ab}/J$. 

However, the renormalization procedure for vertex functions is made difficult
because of complicated expressions (\ref{g_ab})-(\ref{c_d}) for the
operations with the matrices $g_{ab}$. The step-like structure of the 
function $g(x)$ revealed in Refs.~\cite{6,7,8} gives the 
possibility to realize the 
renormalization procedure. In this paper we considered only the matrices 
$g_{ab}$ which have the structure known as the 1 step RSB with function 
$g(x)$ of the next view:
\begin{equation}
 g(x)= \left\{ \begin{array}{c}
  g_0, \quad 0 \leq x < x_0, \\ \displaystyle
  g_1,  \quad x_0 < x \leq 1,
 \end{array} \right.
\end{equation}
where $0\leq x_0 \leq 1$ is the coordinate of the step and it remains
arbitrary parameter. The value of $x_0$ does not be changed during the
renormalization procedure and remains the same as in the starting function
$g_0(x)$. In consequence the RG transformations of the effective replicated
Hamiltonian with RSB potentials are determined by the three parameters
$\tilde{g}$, $g_0$ and $g_1$. 

The critical properties of model can be  extracted  from  the  coefficients
$\beta_i(\tilde{g}, g_0, g_1)$ $(i=1,2,3)$, 
$\gamma_{\phi}(\tilde{g}, g_0, g_1)$, 
and $\gamma_{\phi^2}(\tilde{g}, g_0, g_1)$ 
of  the Callan-Symanzik RG equation. We obtained the $\beta$ functions in the
two-loop approximation in the form of the expansion series in renormalized
parameters $\tilde{g}$, $g_0$ and $g_1$
\begin{eqnarray} \label{beta}
\displaystyle \beta_1&=&-\tilde{g}+\left (8+p\right ){\tilde{g}}^{2}
-p{\it x_0}\,{g_0}^{2}-p\left (1-{\it x_0}\right ){g_1}^{2}+((8\,f
-40\,h+20)p \nonumber \\
\displaystyle &+&16\,f-176\,h+88){\tilde{g}}^{3}+\left (24\,h-8\,f-12
\right ){\it x_0}\,p\tilde{g}{g_0}^{2}+(24\,h-8\,f \nonumber \\
\displaystyle &-&12)\left (1-{\it x_0}\right )p\tilde{g}{g_1}^{2}+\left (16\,h-8\right )
{\it x_0}\,p{g_0}^{3}+\left (16\,h-8\right )\left (1-{\it x_0}
\right )p{g_1}^{3}, \nonumber \\
 \displaystyle\beta_2&=&-g_0+\left (4+2\,p\right )\tilde{g}g_0-
\left (2\,p{\it x_0}-4\right ){g_0}^{2}-2\,\left (1-{\it x_0}\right )pg_0g_1 \nonumber \\
\displaystyle &+&((8\,f-48\,h+28)p+16\,f-48\,h+24){\tilde{g}}^{2}g_0
+(((32\,h-16){\it x_0} \\
\displaystyle &+&8-32\,h )p+48-96\,h )\tilde{g}{g_0}^{2}
+\left (32\,h-16\right)\left (1-{\it x_0}\right )p\tilde{g}g_0g_1 \nonumber \\
\displaystyle &+&(\left (48\,h-8\,f-20\right ){\it x_0}\,p
-32\,h+16){g_0}^{3}+\left (32\,h-8\right )\left (1-{\it x_0}\right )p{g_0}^{2}g_1  \nonumber \\
\displaystyle &+&\left (16\,h-12-8\,f\right )\left (1-{\it x_0}\right )pg_0{g_1}^{2},  \nonumber \\
 \displaystyle\beta_3&=&-g_1-p{\it x_0}\,{g_0}^{2}+\left (p\left
({\it x_0}-2\right )+4\right ){g_1}^{2}+\left (4+2\,p\right )\tilde{g}g_1
+((8\,f-48\,h \nonumber \\
\displaystyle &+&28 )p+16\,f-48\,h+24)g_1{\tilde{g}}^{2}+\left (16\,h-8\right )
{\it x_0}\,p\tilde{g}{g_0}^{2}+( (\left (8-16\,h\right ) {\it x_0} \nonumber \\
\displaystyle &-&8 )p+48-96\,h)u_{{0
}}{g_1}^{2}+\left (16\,h-8\right ){\it x_0}\,p{g_0}^{3}+\left
(8\,h-8\,f-4\right ){\it x_0}\,pg_1{g_0}^{2}\nonumber \\
\displaystyle &+&\left (\left (8\,f-24\,h+12\right ){\it x_0}\,p+\left (48\,h-8\,f
-20\right )p+16-32\,h\right ){g_1}^{3}.\nonumber 
\end{eqnarray}
By analogy with papers \cite{6,7,8} we changed
$g_{a\neq b}\rightarrow - g_{a\neq b}$ in the expressions (\ref{beta}) for the
$\beta$ functions, so that the off-diagonal elements $g_{a\neq b}$ would
be positively defined.

It is well known that perturbation series are asymptotic series, and that
the vertices describing the interaction of the order parameter fluctuations
in the fluctuating region $\tau \rightarrow 0$ are large enough so
that expressions (\ref{beta}) cannot be used directly. For this reason, to
extract the required physical information from the obtained expressions, we
employed the Pad\'{e}-Borel approximation of the summation of asymptotic
series extended to the multiparameter case. The direct and inverse Borel
transformations for the multiparameter case have the form
\begin{eqnarray}
  \displaystyle f(\tilde{g},g_0,g_1)&=&\sum\limits_{i,j,k}c_{ijk}
     \tilde{g}^i g_0^j g_1^k=\int\limits_{0}^{\infty}e^{-t}F(\tilde{g}t,
     g_0t,g_1t)d t,  \\
  \displaystyle F(\tilde{g},g_0,g_1)&=&\sum\limits_{i,j,k}\frac{c_{ijk}}
  {(i+j+k)!}\tilde{g}^i g_0^j g_1^k. 
\end{eqnarray}
A series in the auxiliary variable $\theta$ is introduced for analytical
continuation of the Borel transform of the function:
\begin{equation}
   {\tilde{F}}(\tilde{g},g_0,g_1,\theta)=\sum\limits_{k=0}^{\infty}
   \theta^k\sum\limits_{i=0}^{k}\sum\limits_{j=0}^{k-i}
   \frac{c_{i,j,k-i-j}}{k!}\tilde{g}^i g_0^j g_1^{k-i-j},
\end{equation}
to which the [L/M] Pad\'{e} approximation is applied at the point $\theta=1$.
To perform the analytical continuation, the Pad\'{e} approximant of [L/1] type
may be used which is known to provide rather good results for various
Landau-Wilson models (see, e.g. Refs.~\cite{13,14}). The property of
preserving the symmetry of a system during application of the Pad\'{e}
approximation by the $\theta$ method, as in \cite{13}, has become
important for multivertices models. We used the [2/1] approximant
to calculate the $\beta$ functions in the two-loop approximation.

The nature of the critical behavior is determined by the existence of a
stable FP satisfying the system of equations
\begin{equation}
\beta_{i}(\tilde{g}^*,g_0^*,g_1^*)=0 \quad (i=1,2,3)
\end{equation}
for resummed $\beta$ functions.
We have found three types of non-trivial FP's in the physical region of
parameter space $\tilde{g}^*,g_0^*,g_1^* \geq 0$ for different values of
$p=1,2,3$, which are presented in Tables~\ref{tab1}-\ref{tab4} 
(the exception was made in the
case with $p=3$, when presented in the Table~\ref{tab3} 
the coordinates of type II
and type III FP's are characterized by unphysical negative values of
$g_0^*$ and $g_1^*$). Type I with $\tilde{g}^*\neq 0, g_0^* = g_1^* = 0$
corresponds to the RS FP of a pure system, type II with
$\tilde{g}^*\neq 0, g_0^* = g_1^* \neq 0$ is a disorder-induced RS FP,
and type III with $\tilde{g}^*\neq 0, g_0^* = 0, g_1^* \neq 0$ corresponds to
the 1-step RSB FP. The values of parameters $\tilde{g}^*, g_1^*$ for the
1-step RSB FP depend on the coordinate of the step $x_0$, and we presented
in Tables \ref{tab1}-\ref{tab4}  the received values of these parameters in 
the range $0\leq x_0 \leq 1$ with changes through the step $\Delta x_0 =0.1$.

The type of critical behavior of this disordered system for each value
of $p$ is determined by the stability of the corresponding FP.
The requirement that the FP be stable reduces to the condition that
the eigenvalues $\lambda_i$ of the matrix
\begin{equation}
B_{i,j}=\frac{\partial\beta_i(\tilde{g}^*,g_0^*,g_1^*)}{\partial g_j}
\end{equation}
lie in the right-hand side complex half-plane. Analysis of the received
values $\lambda_i$ for FP's in Tables \ref{tab1}-\ref{tab4} 
shows that for 3D and 2D Ising
models ($p=1$) and 3D XY model ($p=2$) the disorder-induced
RS FP's are stable. However, we believe that in the higher field-theory
orders of approximation the RS FP of a pure system will be stable for the
3D XY model. Two facts indicate this, such as the weak stability of the
disorder-induced RS FP and that in the two-loop approximation the marginal
value of $p_c= 2.0114$ for the borderline between regions of stability for
the disorder-induced RS FP and the RS FP of a pure system. In the higher
orders of approximation the marginal value of $p_c<2$ such as the specific
heat exponent $\alpha >0$ for the pure XY model. For the 3D Heisenberg model
($p=3$) the RS FP of a pure system is stable and both another types of
FP's are characterized by unphysical values of coordinates $g_0^*$ and
$g_1^*$. Our conclusions coincide with results of Ref.~\cite{9} for the 2D 
random Ising model, but contradict with results of Refs.~\cite{6,7,8} for 
3D disordered systems.

However, we must note that the obtained RS FP values for vertices 
$\tilde{g}$, $g_0$ and the eigenvalues $\lambda_1$ and $\lambda_2$ 
of the stability matrix correspond to results of Ref.~\cite{15}, 
in which a field-theoretic treatment of disordered 3D and 2D spin 
systems was presented in the two-loop approximation. The vertices 
$v_1$ and $v_2$ introduced in Ref.~\cite{15} are connected with the 
vertices $\tilde{g}$ and $g_0$  by the relations 
$v_1=(p+8)(\tilde{g}+g_0)$ and $v_2=8g_0$. We have calculated the static
critical exponents from the resummed by the generalized Pad\'{e}-Borel
method $\gamma$ functions in the corresponding stable RS FP's 
(Table~\ref{tab5}).
For comparison we presented also in Table~\ref{tab5} a values of the critical
exponents from \cite{16,17} received for pure and disordered 3D systems
without RSB in the six-loop approximation. Comparison of the exponent
values shows that their differences are not more than 0.01. It gives the
possibility to consider our results of the RSB effects investigation as
reliable. The model with RSB potentials is the another one example of the
multivertices models \cite{13} for which the predictions made on the basis
of the $\varepsilon$ - expansion can differ strongly from results of the use
a more accurate field-theory approach for the 3D system directly together
with methods of series summation. This situation is explained by
a competition of numerous types of FP's in the parameter space of the
multivertices models. Therefore, the spread of results of the
$\varepsilon$ - expansion from $\varepsilon \ll 1$ to $\varepsilon = 1$ is
impossible, as a rule, without intersection of the stability ranges for
the various types of FP's.

Thus, the RG investigations carried out in the two-loop approximation
show the stability of the critical behaviour of weakly disordered systems
with respect to the RSB effects. In dilute Ising-like systems the
disorder-induced critical behaviour is realized with RS FP. The weak
disorder is irrelevant for the critical behaviour of systems with
multicomponent order parameter although the proof of it for 3D systems with
two-component order parameter demands a calculations in the higher
orders of approximation. The possible influence the RSB degrees of
freedom on the critical behaviour of highly disordered systems can be
nonperturbatively revealed by the use of the Monte Carlo simulation method
\cite{18} for definition of the probability distributions for order parameter
and random transition temperature fluctuations. 

\acknowledgments 

We would like to thank the Russian Foundation for Basic
Research for support through Grant No. 00-02-16455.

\widetext

\begin{table}
\squeezetable
\caption{Coordinates of the FP's and eigenvalues of the stability matrix
for the 3D Ising model.}
\label{tab1}
\begin{tabular}{dddddcd}
  Type &$x_0$& $\tilde{g}^*$ & $g_{0}^*$   & $g_{1}^*$     & $\lambda_1 \quad \quad \quad\lambda_2$  & $\lambda_3$ \\  \hline
   1  &     &  0.1774103    &      0      &      0         & $0.65355 \quad  -0.16924$    & $-$0.16924   \\
   2  &     &  0.1843726    &  0.0812240  &   0.0812240    & $0.525\pm0.089i$            &  0.211       \\
   3  & 0.0 &  0.1843726    &      0      &   0.0812240    & $0.525319\pm0.089273 i$      &   $-$0.039167   \\
      & 0.1 &  0.1839722    &      0      &   0.0829404    & $0.535185\pm0.098291 i$      &   $-$0.049185   \\
      & 0.2 &  0.1835134    &      0      &   0.0846432    & $0.547065\pm0.106665 i$      &   $-$0.059851   \\
      & 0.3 &  0.1829917    &      0      &   0.0863186    & $0.560666\pm0.113305 i$      &   $-$0.071187   \\
      & 0.4 &  0.1824035    &      0      &   0.0879503    & $0.576473\pm0.118038 i$      &   $-$0.083210   \\
      & 0.5 &  0.1817458    &      0      &   0.0895200    & $0.595060\pm0.120271 i$      &   $-$0.095927   \\
      & 0.6 &  0.1810165    &      0      &   0.0910067    & $0.617241\pm0.118872 i$      &   $-$0.109334   \\
      & 0.7 &  0.1802154    &      0      &   0.0923872    & $0.643936\pm0.111389 i$      &   $-$0.123415   \\
      & 0.8 &  0.1793442    &      0      &   0.0936384    & $0.675972\pm0.092079 i$      &   $-$0.138133   \\
      & 0.9 &  0.1784070    &      0      &   0.0947426    & $0.713456\pm0.035266 i$      &   $-$0.153431   \\
      & 1.0 &  0.1774103    &      0      &   0.0956920    & $0.857325\quad 0.65355$      &   $-$0.169237  \\
\end{tabular}
\end{table}

\begin{table}
\squeezetable
\caption{Coordinates of the FP's and eigenvalues of the stability
matrix for the 3D XY model.}
\label{tab2}
\begin{tabular}{dddddddd}
  Type &$x_0$& $\tilde{g}^*$ & $g_{0}^*$   & $g_{1}^*$     & $\lambda_1$  &  $\lambda_2$  & $\lambda_3$ \\  \hline
   1  &     & 0.1558303     &    0        &    0          & 0.667315   &$-$0.001672& $-$0.001672  \\
   2  &     & 0.1558310     & 0.0005837   & 0.0005837   & 0.667312   &0.001682     & 0.000004     \\
   3  & 0.0 & 0.1558310     &    0        & 0.0005837   & 0.667313   &0.001683     & $-$0.000001   \\
      & 0.1 & 0.1558310     &    0        & 0.0006143   & 0.667313   &0.001684     & $-$0.000088   \\
      & 0.2 & 0.1558310     &    0        & 0.0006483   & 0.667313   &0.001685     & $-$0.000186   \\
      & 0.3 & 0.1558310     &    0        & 0.0006863   & 0.667313   &0.001686     & $-$0.000296   \\
      & 0.4 & 0.1558310     &    0        & 0.0007291   & 0.667313   &0.001687     & $-$0.000419   \\
      & 0.5 & 0.1558310     &    0        & 0.0007775   & 0.667313   &0.001687     & $-$0.000559   \\
      & 0.6 & 0.1558309     &    0        & 0.0008327   & 0.667313   &0.001688     & $-$0.000717   \\
      & 0.7 & 0.1558308     &    0        & 0.0008964   & 0.667314   &0.001690     & $-$0.000901   \\
      & 0.8 & 0.1558307     &    0        & 0.0009707   & 0.667314   &0.001692     & $-$0.001116   \\
      & 0.9 & 0.1558306     &    0        & 0.0010583   & 0.667315   &0.001694     & $-$0.001369   \\
      & 1.0 & 0.1558303     &    0        & 0.0011633   & 0.667316   &0.001696     & $-$0.001672   \\
\end{tabular}
\end{table}

\begin{table}
\squeezetable
\caption{Coordinates of the FP's and eigenvalues of the stability
matrix for the 3D Heisenberg model.}
\label{tab3}
\begin{tabular}{dddddddd}
  Type &$x_0$& $\tilde{g}^*$ & $g_{0}^*$   & $g_{1}^*$     & $\lambda_1$  &  $\lambda_2$  & $\lambda_3$ \\  \hline
   1  &     & 0.1382700     &   0          &     0          & 0.681378   &  0.131537  & 0.131537     \\
   2  &     & 0.1419323     &$-$0.0358629  & $-$0.0358629   & 0.672676   &$-$0.089135 & $-$0.005783  \\
   3  & 0.0 & 0.1419323     &   0          & $-$0.0358629   & 0.672676   &$-$0.089135 & $-$0.005783  \\
      & 0.1 & 0.1419931     &   0          & $-$0.0381865   & 0.672729   &$-$0.086515 & 0.001104     \\
      & 0.2 & 0.1420386     &   0          & $-$0.0408334   & 0.672845   &$-$0.083560 & 0.008802     \\
      & 0.3 & 0.1420600     &   0          & $-$0.0438761   & 0.673046   &$-$0.080206 & 0.017469     \\
      & 0.4 & 0.1420441     &   0          & $-$0.0474104   & 0.673361   &$-$0.076366 & 0.027299     \\
      & 0.5 & 0.1419699     &   0          & $-$0.0515650   & 0.673831   &$-$0.071931 & 0.038540     \\
      & 0.6 & 0.1418040     &   0          & $-$0.0565177   & 0.674509   &$-$0.066754 & 0.051518     \\
      & 0.7 & 0.1414913     &   0          & $-$0.0625193   & 0.675476   &$-$0.060636 & 0.066656     \\
      & 0.8 & 0.1409374     &   0          & $-$0.0699349   & 0.676837   &$-$0.053299 & 0.084517     \\
      & 0.9 & 0.1399720     &   0          & $-$0.0793131   & 0.678742   &$-$0.044336 & 0.105835     \\
      & 1.0 & 0.1382700     &   0          & $-$0.0915089   & 0.681378   &$-$0.033124 & 0.131537     \\
\end{tabular}
\end{table}

\begin{table}
\squeezetable
\caption{Coordinates of the FP's and eigenvalues of the stability matrix
for the 2D Ising model.} 
\label{tab4}
\begin{tabular}{dddddddd}
  Type &$x_0$& $\tilde{g}^*$ & $g_{0}^*$   & $g_{1}^*$     & $\lambda_1$  &  $\lambda_2$  & $\lambda_3$ \\  \hline
   1  &     &  0.2693889    &      0      &      0         & 0.461180       &  $-$0.095180        &  $-$0.095180   \\  \hline
   2  &     &  0.2735768    &$-$0.0645918 & $-$0.0645918   & 0.446348       &  0.162092           &     0.056959    \\  \hline
   3  & 0.0 &  0.2735768    &      0      & $-$0.0645918   & 0.446348       &  0.162092           &  $-$0.025112   \\
      & 0.1 &  0.2733970    &      0      & $-$0.0665073   & 0.446636       &  0.165301           &  $-$0.029995   \\
      & 0.2 &  0.2731825    &      0      & $-$0.0685156   & 0.447048       &  0.168803           &  $-$0.035240   \\
      & 0.3 &  0.2729279    &      0      & $-$0.0706204   & 0.447615       &  0.172637           &  $-$0.040881   \\
      & 0.4 &  0.2726271    &      0      & $-$0.0728248   & 0.448377       &  0.176842           &  $-$0.046958   \\
      & 0.5 &  0.2722731    &      0      & $-$0.0751310   & 0.449385       &  0.181468           &  $-$0.053513   \\
      & 0.6 &  0.2718579    &      0      & $-$0.0775405   & 0.450704       &  0.186566           &  $-$0.060592   \\
      & 0.7 &  0.2713724    &      0      & $-$0.0800528   & 0.452416       &  0.192194           &  $-$0.068245   \\
      & 0.8 &  0.2708067    &      0      & $-$0.0826660   & 0.454631       &  0.198417           &  $-$0.076523   \\
      & 0.9 &  0.2701495    &      0      & $-$0.0853752   & 0.457490       &  0.205298           &  $-$0.085483   \\
      & 1.0 &  0.2693890    &      0      & $-$0.0881727   & 0.461180       &  0.212903           &  $-$0.095180   \\   \hline
\end{tabular}
\end{table}

\begin{table}
\squeezetable
\caption{Critical exponents of the 3D models for RS FP's }
\label{tab5}
\begin{tabular}{lllddddd}
    Model   &  FP  &                          &  $\eta$   &   $\nu$   & $\gamma$  & $\beta$  & $\alpha$      \\ \hline
    Ising   &  RS1 &  this work               & 0.028     & 0.631     & 1.244     & 0.324    &    0.107      \\
            &      &  Ref.~\protect{\cite{16}}& 0.031(4)  & 0.630(2)  & 1.241(2)  & 0.325(2) &    0.110(5)   \\
            &  RS2 &  this work               & 0.028     & 0.672     & 1.329     & 0.345    & $-$0.015      \\
            &      &  Ref.~\protect{\cite{17}}& 0.030(3)  & 0.678(10) & 1.330(17) & 0.349(5) & $-$0.034(30)  \\ \hline
       XY   &  RS1 &  this work               & 0.029     & 0.667     & 1.318     & 0.343    & $-$0.001      \\
            &      &  Ref.~\protect{\cite{16}}& 0.034(3)  & 0.669(1)  & 1.316(1)  & 0.346(1) & $-$0.007(6)   \\ \hline
Heisenberg  &  RS1 &  this work               & 0.028     & 0.697     & 1.379     & 0.369    & $-$0.092      \\
            &      &  Ref.~\protect{\cite{16}}& 0.034(3)  & 0.705(1)  & 1.387(1)  & 0.364(1) & $-$0.115(9)   \\
\end{tabular}
\end{table}

\end{document}